\documentclass[12pt]{article}
\usepackage{amsmath, amsthm, amsfonts}
\usepackage{amssymb,mathrsfs}
\usepackage{float, multirow, url}
\usepackage[margin=1in]{geometry}
\usepackage{graphicx}
\usepackage{tabularx, booktabs}
\usepackage{enumerate}
\usepackage{color}
\usepackage[dvipsnames]{xcolor}
\usepackage{natbib,afterpage}

\newcommand{\blind}{1}

\def\T{{\mathrm{\scriptscriptstyle T}}}

\newcommand{\nnskip}{\nonumber \\}

\newcommand{\floor}[1]{\lfloor #1 \rfloor}
\newcommand{\abs}[1]{\lvert #1 \rvert}
\newcommand{\cov}{\textup{cov}}

\def\T{{\mathrm{\scriptscriptstyle T}}}
\def\tr{\mbox{tr}}

\newtheorem{theorem}{Theorem}
\newtheorem{corollary}{Corollary}
\newtheorem{lemma}{Lemma}
\newtheorem{condition}{Condition}

\def\tr{\mbox{tr}}

\def\bi{I}
\def\1{\mathbf{1}}

\def\bY{Y}

\def\bsig{\Sigma}

\def\tr{\mbox{tr}}

\def\bsc{\begin{scriptsize}}
\def\esc{\end{scriptsize}}

\def\be{\begin{equation}}
\def\ee{\end{equation}}
\def\bea{\begin{eqnarray}}
\def\eea{\end{eqnarray}}
\def\bd{\begin{displaymath}}
\def\ed{\end{displaymath}}
\def\bda{\begin{eqnarray*}}
\def\eda{\end{eqnarray*}}
\def\bsm{\begin{small}}
\def\esm{\end{small}}
\begin{document}

\def\spacingset#1{\renewcommand{\baselinestretch}%
{#1}\small\normalsize} \spacingset{1}

%%%%%%%%%%%%%%%%%%%%%%%%%%%%%%%%%%%%%%%%%%%%%%%%%%%%%%%%%%%%%%%%%%%%%%%%%%%%%%

\if1\blind
{
  \title{Homogeneity Tests of Covariance and Change-Points Identification for High-Dimensional Functional Data}
  \centering
  \author{Shawn Santo\thanks{
    Shawn Santo is an Assistant Professor of the Practice, Department of Statistical Science, Duke University, Durham, NC 27708 (e-mail: shawn.santo@duke.edu).}\\
    Department of Statistical Science, Duke University,\\
    and \\
    Ping-Shou Zhong\thanks{
    Ping-Shou Zhong is an Associate Professor, Department of Mathematics, Statistics, and Computer Science, University of Illinois at Chicago, Chicago, IL 60607 (e-mail: pszhong@uic.edu).}\\
    Department of Mathematics, Statistics, and Computer Science,\\ University of Illinois at Chicago}
   \date{}
  \maketitle
} \fi

\if0\blind
{
  \begin{center}
    {\LARGE\bf Title}
\end{center}
} \fi

\vspace{-1cm}
\begin{abstract}
We consider inference problems for high-dimensional functional data with a dense number ($T$) of repeated measurements taken for a large number of $p$ variables from a small number of $n$ experimental units. 
The spatial and temporal dependence, high dimensionality, and the dense number of repeated measurements all make theoretical studies and computation challenging. This paper has two aims; our first aim is to solve the theoretical and computational challenges in detecting and identifying change points among covariance matrices from high-dimensional functional data. The second aim is to provide computationally efficient and tuning-free tools with a guaranteed stochastic error control. The change point detection procedure is developed in the form of testing the homogeneity of covariance matrices. The weak convergence of the stochastic process formed by the test statistics is established under the ``large $p$, large $T$ and small $n$'' setting. Under a mild set of conditions, our change point identification estimator is proven to be consistent for change points at any location of a sequence. Its rate of convergence is shown to depend on the data dimension, sample size, number of repeated measurements, and signal-to-noise ratio. We also show that our proposed computation algorithms can significantly reduce the computation time and are applicable to real-world data with a large number of high-dimensional repeated measurements (e.g. fMRI data). 
Simulation results demonstrate both finite sample performance and computational effectiveness of our proposed procedures.  
We observe that the empirical size of the test is well controlled at the nominal level, and the locations of multiple change points can accurately be identified. An application to fMRI data demonstrates that our proposed methods can identify event boundaries in the preface of the movie {\it Sherlock}. Our proposed procedures are implemented in an R package \emph{TechPhD}.
\end{abstract}

\noindent%
\vfill

\newpage
\spacingset{1.5} % DON'T change the spacing!

{\centering{
\section{Introduction}
\label{sec:introduction} }}

The covariance matrix or precision matrix, defined as the inverse of the 
covariance matrix, is very commonly used in applications to characterize 
dependence. For example, in fMRI studies the covariance matrix is
often used to quantify relationships between spatially separated brain 
regions. Neuroscientists and statisticians use these relationships to better understand
the brain's functional connectivity. Our work is motivated by an experiment conducted by Chen et al. (2017), where 17 participants watched
the same 48-minute segment of the BBC television series {\it Sherlock}.
The study cultivated blood-oxygen-level dependent (BOLD) readings 
in spatially separated brain regions for all subjects. The goal of the experiment was to understand perception and memory processes in 
humans as they experience continuous real-world events, such as watching a movie. 
Event segmentation theory, posited by Zacks et al. (2007), suggests that humans 
may partition a continuous experience into a series of segmented discrete events
for memory storage. Schapiro et al. (2013) has suggested that these event 
boundaries from the partitions are created around changes in functional 
connectivity. Identifying change points with regards to functional connectivity via the covariance matrix
provides an objective and meaningful approach to discover event boundaries. This fosters our understand of
how subjects processed the continuous stream of audio and visual stimuli exposed to them during their
viewing of {\it Sherlock}.

Assume $Y_{it}=(Y_{it1},\ldots, Y_{itp})^\T$ is a $p$-dimensional random vector with mean vector $\mu_t$ and covariance matrix 
$\Sigma_t$. In a typical fMRI study at the voxel level, $Y_{it} \ (i = 1, \dots, n; \ t=1, \ldots, T)$ represents the $p$ BOLD signal 
measurements for the $i$-th individual observed at the $t$-th time point, where
$p$, $T$, and $n$ are at the order of 100,000, 100, and 10, respectively. 
For the region of interest (ROI) analysis, $p$ represents the number of 
spatially separated brain regions and is at the order of 100. 
We are dealing with a high dimensional functional data setting where the dimension of functional curves ($p$)
are comparable or far exceeds $T$ and $n$. This setting is different from the classical multivariate functional data setting where
$p$ is fixed and does not grow with $n$ and $T$. (e.g., Chiou et al, 2014). 
The goal of this paper is to develop nonparametric, computationally 
efficient and tuning-free statistical procedures to detect and identify change 
points among covariance matrices in high-dimensional functional data with a dense number 
of repeated measurements (Li and Hsing, 2010). 
The covariance change point detection problem can be posed in the form of a statistical hypothesis 
test as
\begin{align}
\label{hyp-infinite-T}
H_0: & \ \Sigma_1 = \cdots = \Sigma_T=\Sigma \qquad \mbox{versus} \nnskip
H_1: & \ \Sigma_1 = \cdots = \Sigma_{\tau_1} \neq \Sigma_{\tau_1 + 1}
                  = \cdots = \Sigma_{\tau_q} \neq \Sigma_{\tau_q + 1}
                  = \cdots = \Sigma_T,
\end{align}
where $\tau_k < T \ (k = 1, \ldots, q < \infty)$ are the unknown change point locations. 
If we reject stationarity among $\Sigma_t$, 
the second task is to estimate the unknown locations of change points $\tau_k$s. 

In functional data literature, $Y_{it}$ is considered as a realization of $p$-dimension functional curves measured with errors observed at time $t$.
Many recently developed inference methods for low-dimensional (small $p$) functional data are summarized in Horv\'{a}th and Kokoszka (2012).  
Most inference methods designed for the low-dimensional cases can not be applied to our setting because 
they depend on functional principal components analysis (FPCA) in the final implementation (Horv\'{a}th and Kokoszka, 2012; Hall et al., 2006). 
FPCA has been demonstrated to not be reliable for high-dimensional functional data. One reason is that the smoothing and PCA steps are very computationally 
expensive, if not infeasible, due to the high dimensionality and the large number of repeated measurements of $Y_{it}$ (Xiao et al., 2016). 
Second, PCA is not consistent when the data dimension $p$ is larger than $n$ (Jung and Marron, 2009). 
To overcome these difficulties, we develop a new statistical inference approach for high-dimensional functional data.
Our approach avoids the smoothing and FPCA steps and provides a computationally efficient and statistically rigorous approach for high-dimensional functional data. 
Although we consider a high-dimensional setting, we do not require a sparsity assumption for $\Sigma_t$, which is commonly used 
in high-dimensional covariance or precision matrix estimation literature (e.g., Bickel and Levina, 2008; Pourahmadi, 2013). In addition, we admit a
general spatiotemporal dependence structure in $\{Y_{it}\}_{t=1}^T$, and stationarity in temporal dependence among $\{Y_{it}\}_{t=1}^T$ is not required. 
Stationarity implies that the temporal dependence between $Y_{it}$ and $Y_{is}$ is the same as that between $Y_{iu}$ and $Y_{iv}$ if $t-s=u-v\neq 0$.

The hypothesis test considered in (\ref{hyp-infinite-T}) is also related but significantly different 
from homogeneity tests of covariance matrices in multivariate 
statistical analysis. For low dimensional data with a fixed data dimension
 $p$, Anderson (2003) and Muirhead (2005) detailed the classical likelihood 
ratio tests. However, the likelihood ratio tests require $n$ to 
exceed $p$ and are only applicable to 
temporally independent samples. The multivariate tests for the homogeneity
of high-dimensional covariance matrices have received much attention in 
the past few years.  A partial list includes Schott (2007), Srivastava and 
Yanagihara (2010), Li and Chen (2012), Zhang et al. (2018) and Ishii et al. (2016, 2019). 
Most of the existing research developed tests under a temporal independence assumption
except for Zhong et al. (2019) who considered high-dimensional longitudinal data with $T$ fixed and small. 
All existing methods considered an asymptotic setting with a fixed number of groups or repeated measurements 
(i.e., $T$ is fixed and small). It is worthwhile to emphasize that extending finite
$T$ results to diverging $T$ is challenging both theoretically and computationally.

Our proposed methods are naturally connected to covariance change point literature. In addition to a change point
detection method, we propose a method to estimate unknown locations of change points. 
Much of the research in high-dimensional covariance change point identification considers the scenario with temporal independent samples. 
For instance, Wang et al. (2017) considered covariance matrix change point identification for $T$ independent 
$p$-dimensional sub-Gaussian random vectors which requires $p < T$. 
Dette et al. (2018) proposed a two-stage covariance change point identification 
procedure based on $T$ independent sub-Gaussian random vectors. In stage
one their procedure involved dimension reduction governed by a regularization 
parameter. In stage two they used a CUSUM-type statistic to estimate the 
locations of change points. Despite these recent advances, none of the 
aforementioned change point identification methods are applicable to high-dimensional functional data.
We study the rate of convergence of the proposed change point estimator under an asymptotic setting that 
is suitable for high-dimensional dense functional data where $p$, $n$, and $T$ all diverge, but we do not require restrictive 
conditions on the relationships between $p, n$ and $T$.  Our
proposed method is able to consistently estimate change points $\tau_k$ that occur
at any location in a sequence such that $\tau_k\asymp T^{\delta_k}$ for some $0\leq \delta_k\leq1$. 
This is more broader than the typical assumption of $\tau_k\asymp T$ in the existing literature.

This paper provides both theoretical and computational contributions. From a 
theoretical perspective, we investigate covariance tests and change point identification for 
high-dimensional functional data, under a setting
in which $n$, $p$, and $T$ diverge. For $T$ diverging, the test statistic 
forms a stochastic process. The convergence of the finite-$T$ distributions (e.g., Zhong et al., 2019)
is not sufficient for weak convergence of a stochastic process with diverging $T$. 
In addition, the complex temporal and spatial dependence must be carefully addressed.
Thus, it is non-trivial to establish weak convergence of our proposed test statistic.
We also discover a parameter-free asymptotic distribution under the temporal stationary assumption, which
makes a connection with the classical literature. 
Furthermore, we derive the rate of convergence for the change point estimator for
change points at any location in the sequence.
Our investigation reveals that the rate of convergence depends on the 
data dimension, sample size, number of repeated measurements, and 
signal-to-noise ratio. The change point identification estimator is shown to 
be consistent, even for the change points that are close to the boundaries,
provided the signal strength exceeds the noise. To the best of our knowledge, 
the asymptotic framework in which $n$, $p$, and $T$ all diverge has not previously 
been investigated with regards to change point identification among high-dimensional covariance matrices.

From a computational perspective, we develop an efficient algorithm 
for our methods, and thus make it practical to apply our procedure to fMRI data 
sets and others with similar structure. We introduce two recursive 
relationships and computation efficient formulae as a way to reduce the 
computation complexity from $O(pn^4T^6)$ to $O(pn^2T^4)$. A quantile approximation 
technique is shown to further decrease the complexity to the order of $pn^2T^3$. 
The approximation technique's accuracy is demonstrated through 
simulation studies. These improvements are included in an R package, 
{\it TechPhD}, an abbreviation of ``Tests and Estimation of Covariance cHange-Points for High-dimensional Data"
which provides an option for parallel computing.

From a practical point of view, our proposed method is attractive because it is free of tuning parameters, 
and has a guaranteed stochastic error control under very general assumptions.
Researchers in neuroscience have developed various methods to study dynamic
functional brain connectivity for single patients and populations. 
Most of the existing work studies dynamic functional connectivity 
using a sliding window approach (e.g., Monti et al., 2014) or 
regularization approach (e.g., Kundu et al., 2018) by directly estimating the 
locations of change points without detecting the existence of change points. 
Both of the aforementioned approaches depend on tuning parameters; these are often difficult to
select in practice. In general, these change point identification methods 
developed in neuroscience are ad hoc and lack the theoretical rigor to 
ensure a consistent and robust procedure. Our proposed procedure is free of 
tuning parameters and is theoretically rigorous. There do exist some 
resampling methods for detecting the existence of change points. However, 
most of these methods assume temporal independence or stationary temporal 
dependence with a specific structure. For example, vector autoregressive 
models are applied in Barnett and Onnela (2016) and Zalesky et al. (2014), 
and strict stationarity was used in the bootstrap procedure proposed 
in Cribben et al. (2013).

The remaining sections of this paper are organized as follows. 
Section 2 details the statistical model and our basic setting. 
Section 3 introduces our proposed test procedures. 
The test statistic's asymptotic distribution is derived under the asymptotic 
framework in which $n$, $p$, and $T$ diverge. Computation consideration with 
regards to the test procedure is provided in Section 4. 
Section 5 introduces an estimator to identify the locations of change points should we reject $H_0$ of 
(\ref{hyp-infinite-T}). The estimator's rate of convergence is studied, and a binary segmentation 
procedure is detailed to estimate the locations of multiple change points. 
Sections 6 and 7 demonstrate the finite sample performance via 
simulation and an application to event segmentation through the 
motivating example, respectively. All proofs and technical details are 
provided in the Supplementary material.

\setcounter{section}{1}
{{\centering
\section*{2. Data model and basic setting}
\label{sec:model}
}}
Suppose we have $n$ independent individuals that have $p$ variables recorded 
at each of $T$ identical time points. Let $\bY_{it}=(\bY_{it1},\ldots,\bY_{itp})^\T$ 
be an observed $p$-dimensional random vector, where 
$Y_{it} \ (i = 1, \dots, n; \ t=1, \ldots, T)$ is independently and identically 
distributed for all $n$ individuals. Assume $Y_{it}$ follows a general factor 
model, where
\begin{align}
\label{model-infinite-T}
Y_{it} = \mu_t + \Gamma_t Z_i,
\end{align}
and $\mu_t$ is a $p$-dimensional unknown mean vector, $\Gamma_t$ is an unknown 
$p \times m$ matrix such that $m \geq pT$, and $Z_i$'s are independent 
$m$-dimensional multivariate standard normal random 
vectors. Since $\textup{var}(Z_i) = I_m$, it follows that for the $i$th 
individual, $\textup{cov}(\Gamma_s Z_i, \Gamma_t Z_i) = \Gamma_s\Gamma_t^{\T}$. 
We define $\Gamma_s\Gamma_t^{\T}$ as $C_{st}$ for different 
time points, $s$ and $t$, and define $\Gamma_t\Gamma_t^{\T}$ as $\Sigma_t$. 
Thus, for the $i$th individual, $\cov(Y_{is}, Y_{it})=C_{st}$ if $s \neq t$ and 
$\cov(Y_{is}, Y_{it})=\Sigma_t$ if $s=t$
for all $s, t \in \{1, \ldots, T\}$. For individuals $i \neq j$, $\cov(Y_{is}, Y_{jt}) = 0$. 
By definition, $C_{st}$ and $\Sigma_t$ are $p \times p$ matrices for all 
$s, t \in \{1, \ldots, T\}$. No specific structure is required on covariance 
matrices $C_{st}$ and $\Sigma_t$. Their generality allows us to capture the 
spatiotemporal dependence in and among the random vectors 
$Y_{it} \ (i = 1, \dots, n; \ t=1, \ldots, T)$. In the context of fMRI data, spatial 
dependence is present among neighboring voxels or nodes and is captured in both 
$C_{st}$ and $\Sigma_t$. Temporal dependence exists for the same voxel or node 
across time points and is captured in matrix $C_{st}$.

The factor model in (\ref{model-infinite-T}) includes the commonly used models 
(e.g., Horv\'{a}th and Kokoszka, 2012; Li and Hsing, 2010) in functional data literature 
as a special case where $Y_{it}$ is assumed to be generated from the following model:
\begin{equation}
\label{functional-model}
Y_{it}=\mu_t+X_i(t)+\varepsilon_{it},
\end{equation}
where $X_i(t)$ is a Gaussian process with mean 0 and covariance function 
$\textup{cov}\{X_i(t), X_i(s)\}=C_{x}(t,s)$ and $\varepsilon_{it}$ are 
independent normally distributed nugget effects with variance 
$\textup{var}(\varepsilon_{it})=\sigma^2 \bi_p$. 
Under the model (\ref{functional-model}), the null hypothesis in (\ref{hyp-infinite-T})
is equivalent to test the equivalence of the covariances $C_{x}(t,t)$ for $t=1,\cdots, T$.
The model (\ref{functional-model}) is a special case of model (\ref{model-infinite-T}),
by defining $\Gamma=(\Gamma_1^{\T},\cdots, \Gamma_T^{\T})^{\T}=(C_x+\sigma^2\bi_{pT})^{1/2}$ 
where $C_x=\{C_x(i, j)\}_{i, j=1}^{\T}$ is a $pT\times pT$-dim matrix. 

%\ccom{[Add a transition before this sentence.] One common approach in functional 
%data analysis is ``smoothing first'' and then 
%applying functional principal components analysis (PCA) to the smoothed 
%curves (Hall et al., 2006).}
%However, this approach is not feasible for high-dimensional functional data. 
%\cre{One reason is that} the smoothing and PCA steps are very computationally 
%expensive (if not infeasible) due to the high dimensionality and the large 
%umber of repeated measurements of $Y_{it}$ (Xiao et al., 2013). 
%Second, the functional PCA is also problematic when the data dimension $p$ is 
%larger than $n$ (Jung and Marron, 2009). \cre{Lastly}, the statistical inference
%after dimension reduction (e.g., PCA) is not reliable if no adjustment is made 
%Kim et al., 2019). \ccom{[Needs some clarity.]The post 
%inference after dimension reduction problem would 
%be more serious under the high-dimensional setting. 
%To overcome these difficulties, we develop a new approach for statistical 
%inference for high-dimensional functional data
%to avoid the smoothing and functional PCA steps.}

\setcounter{section}{2}
{\centering{
\section*{3. Homogeneity tests of covariance matrices}
\label{sec:change-point-detection}
}}

To avoid the inconsistency and uncertainty due to the dimension reduction methods as in functional PCA,
our proposed method directly solves the testing problem by considering a measure to quantify 
the differences among the high-dimensional covariances $\bsig_t$'s. 
We use $D_t$ $(t = 1, \ldots, T-1)$ to measure the averaged distance of 
the covariance matrices before and after time $t$, where 
\begin{align}
\label{Dt-infinite-T}
D_t=\frac{1}{w(t)}\sum_{s_1=1}^t\sum_{s_2=t+1}^T\tr\{(\bsig_{s_1}-\bsig_{s_2})^2\},
\end{align}

\afterpage{%
\newgeometry{left=0.16cm, right=2cm, vmargin=0cm, footnotesep=0.5cm}
% material for this page
\begin{figure}[H]
  \includegraphics{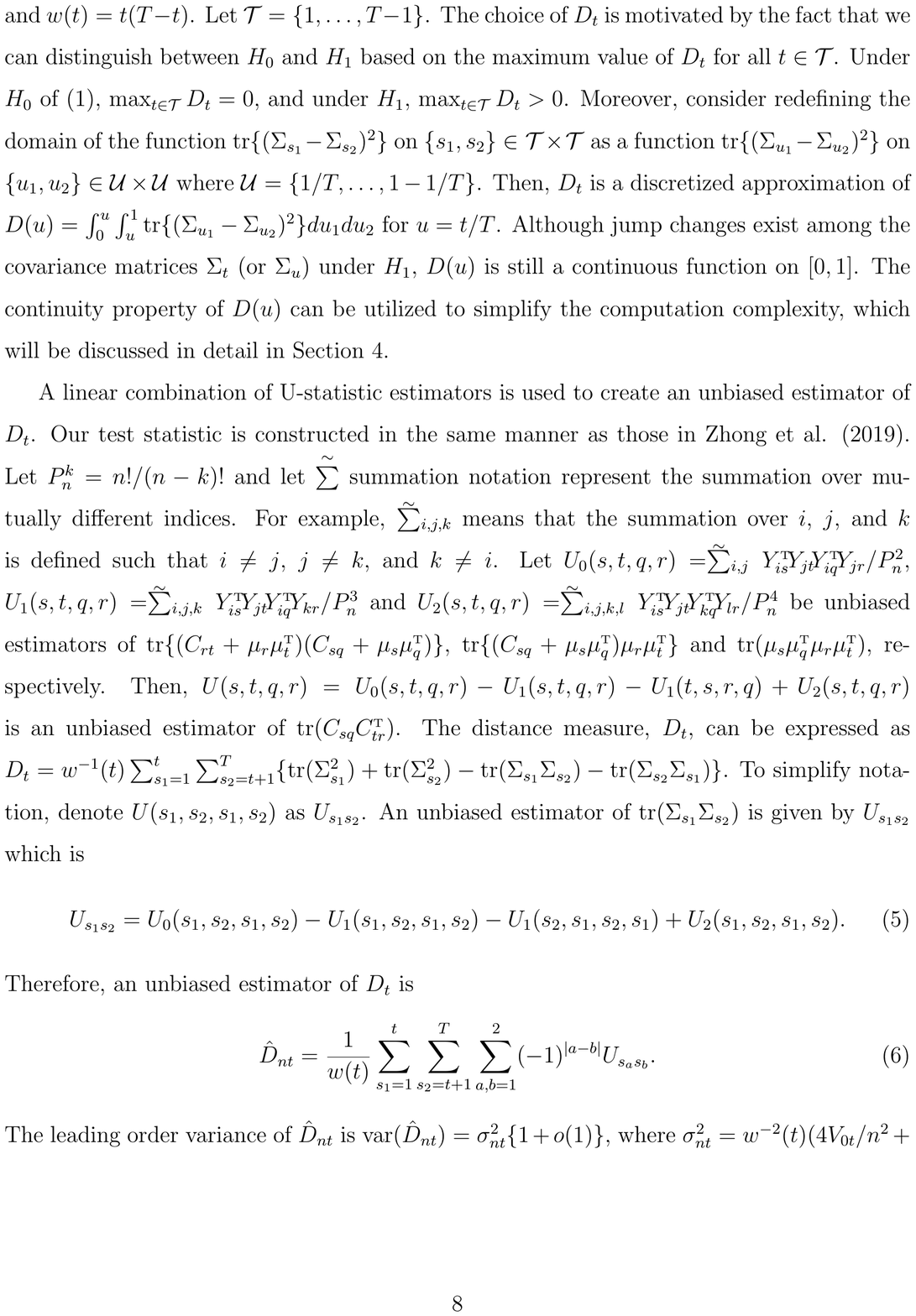}
\end{figure}

\begin{figure}[H]
  \includegraphics{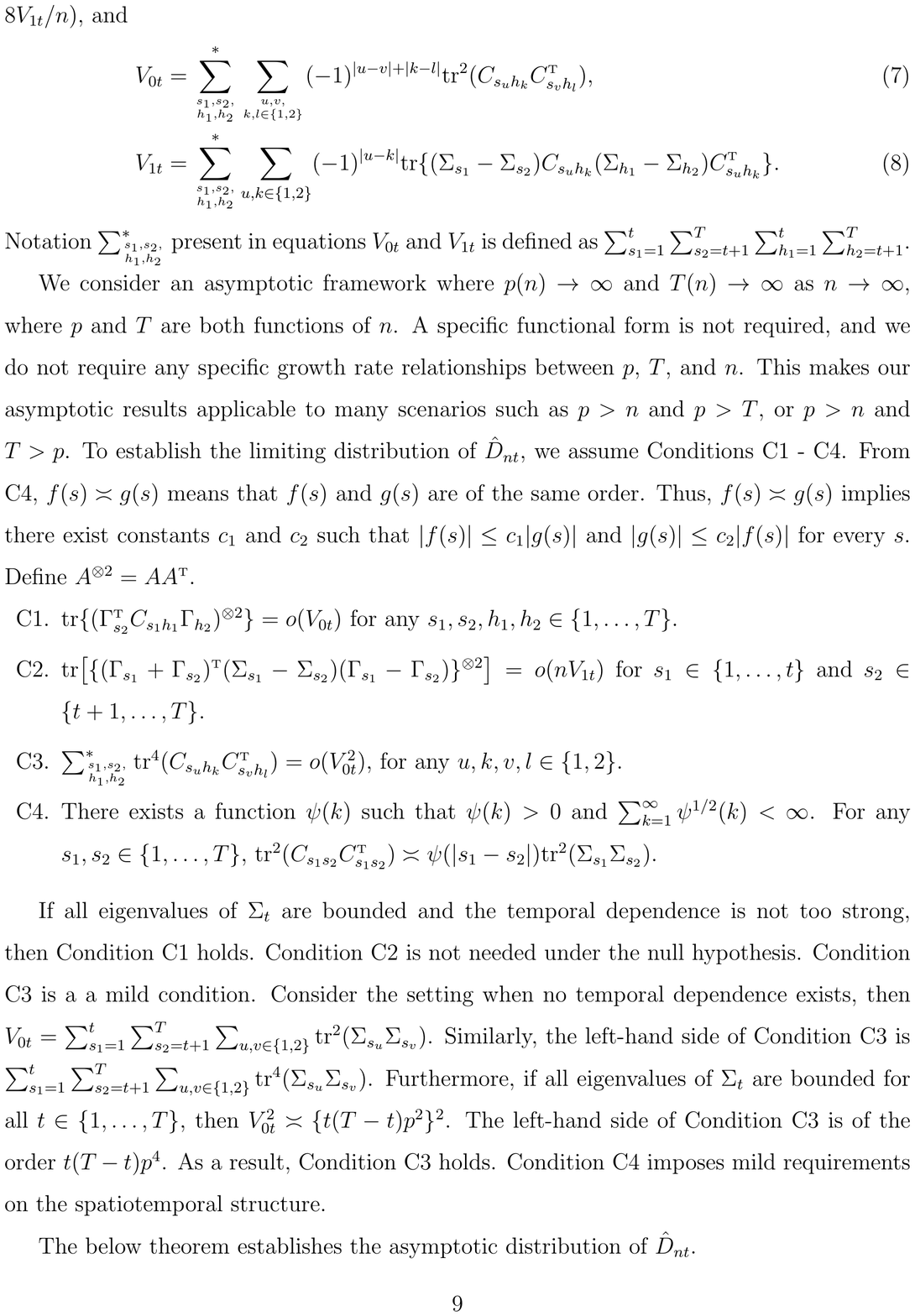}
\end{figure}

\begin{figure}[H]
  \includegraphics{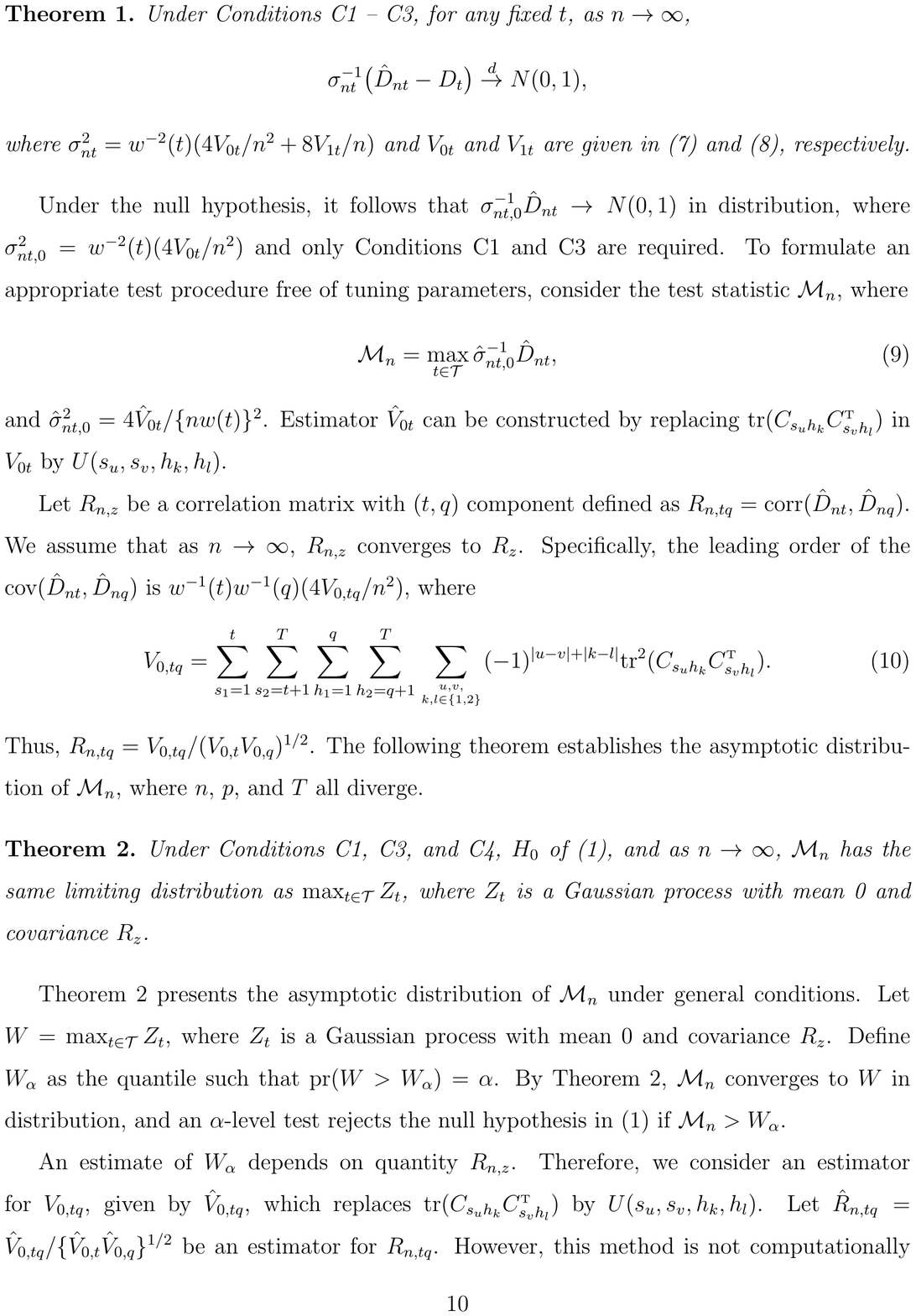}
\end{figure}

\begin{figure}[H]
  \includegraphics{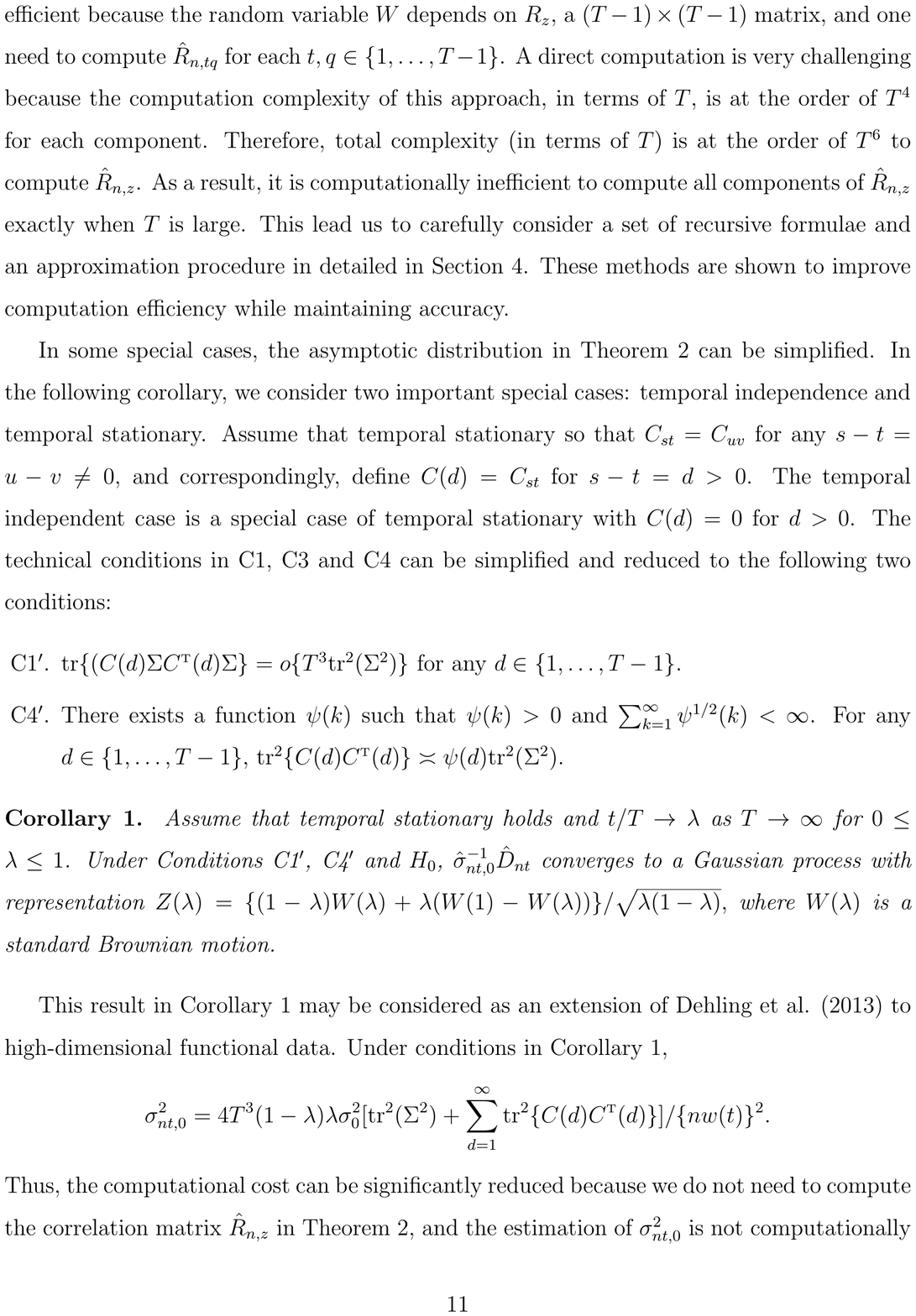}
\end{figure}

\begin{figure}[H]
  \includegraphics{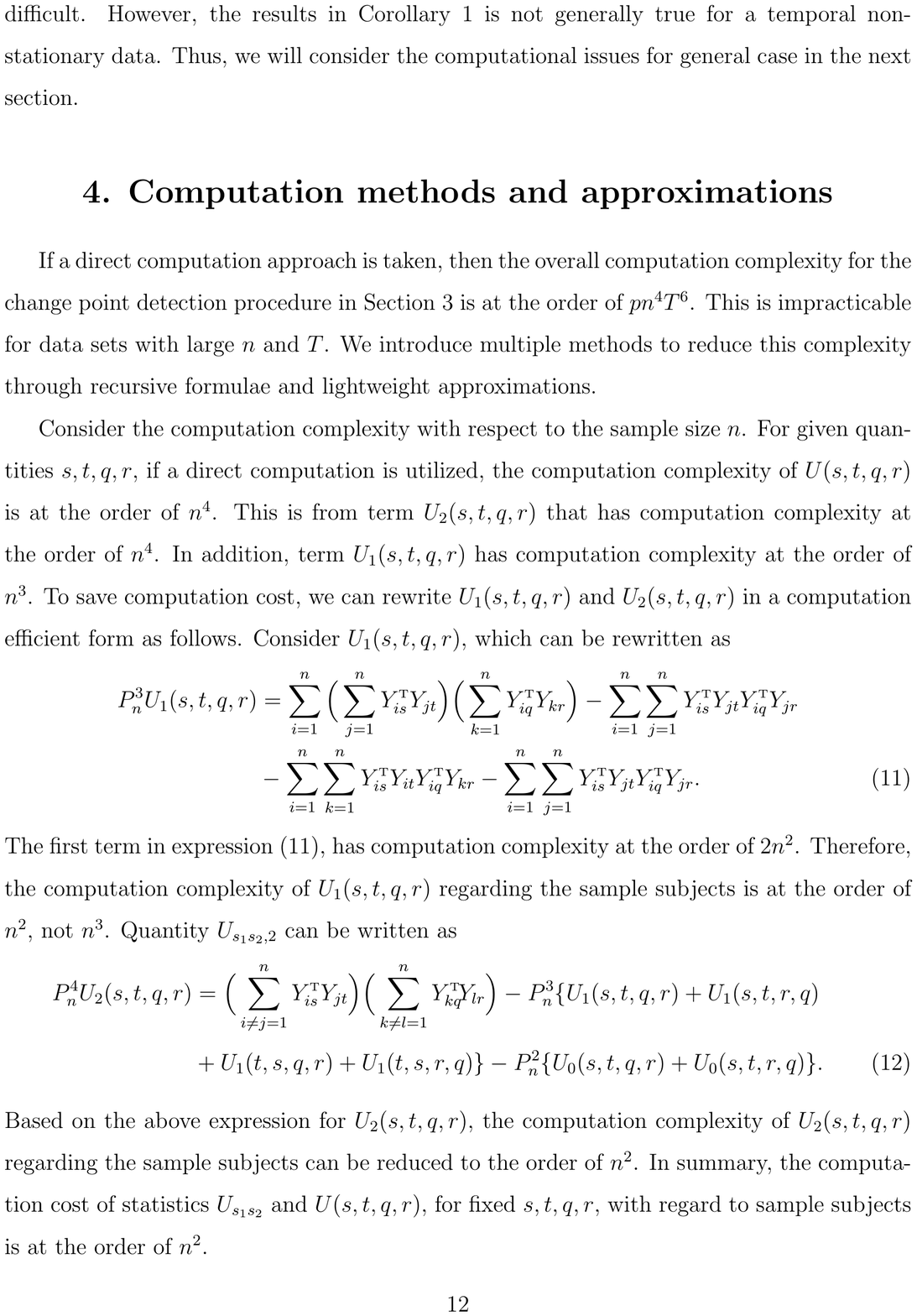}
\end{figure}

\begin{figure}[H]
  \includegraphics{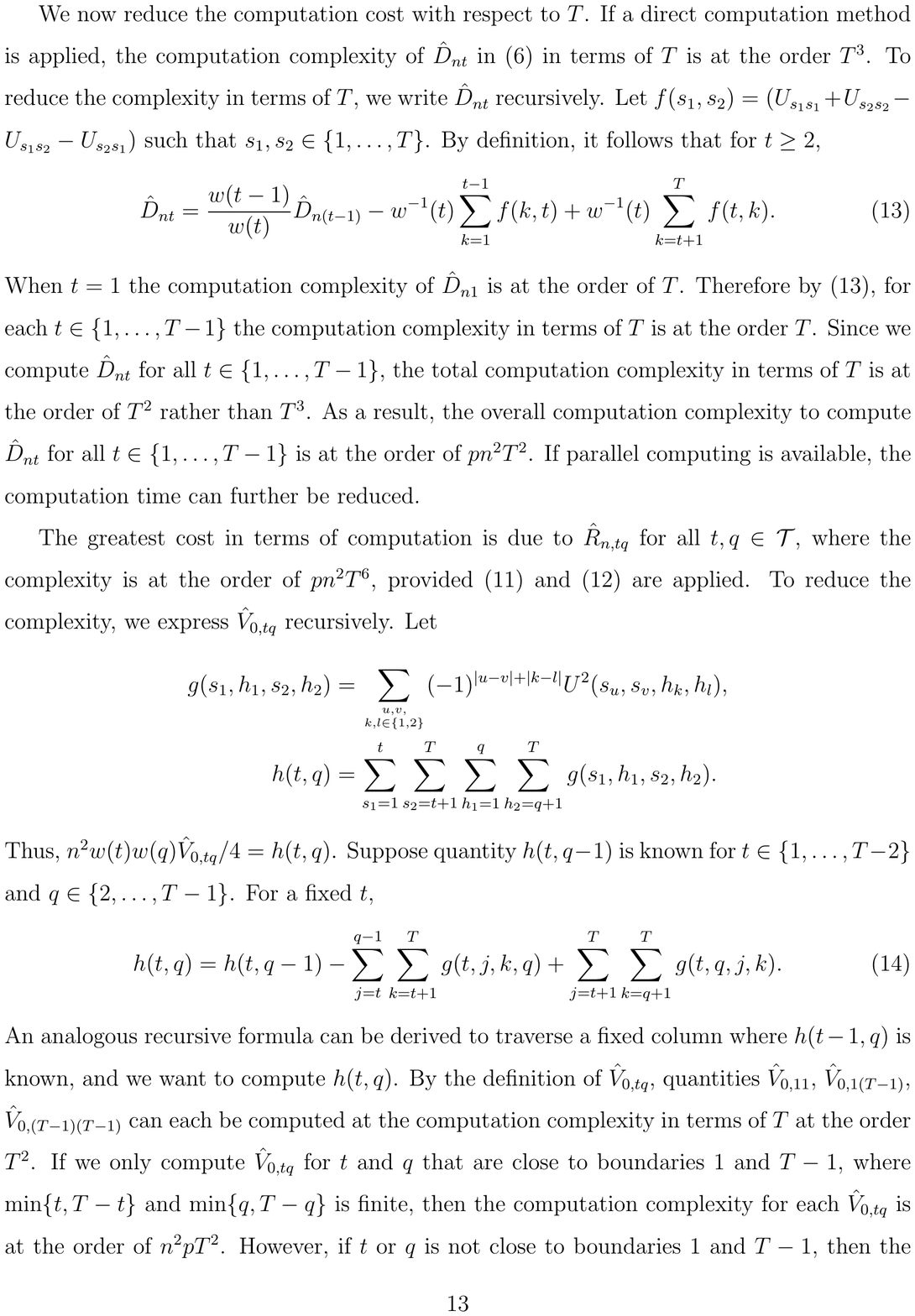}
\end{figure}

\begin{figure}[H]
  \includegraphics{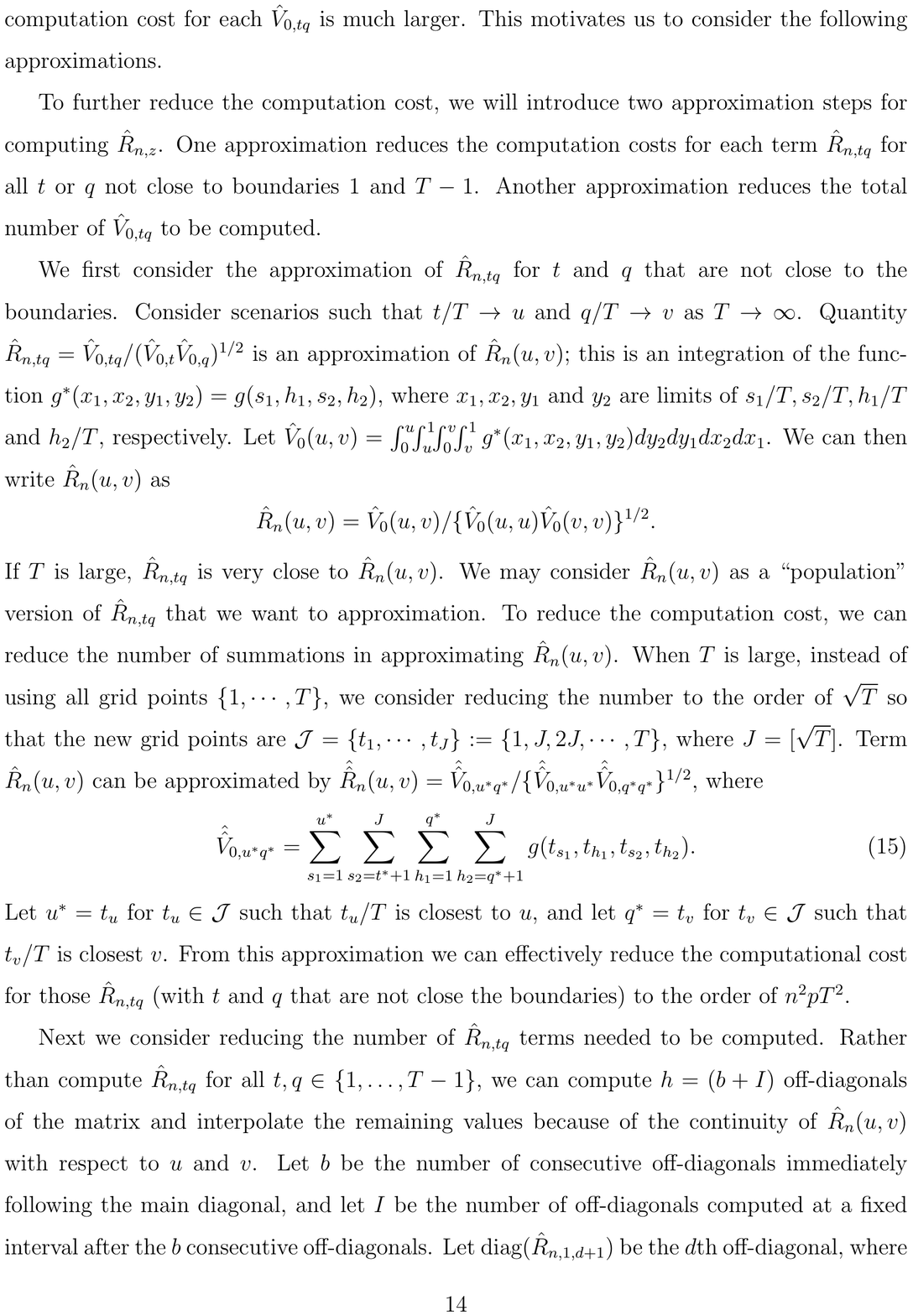}
\end{figure}

\begin{figure}[H]
  \includegraphics{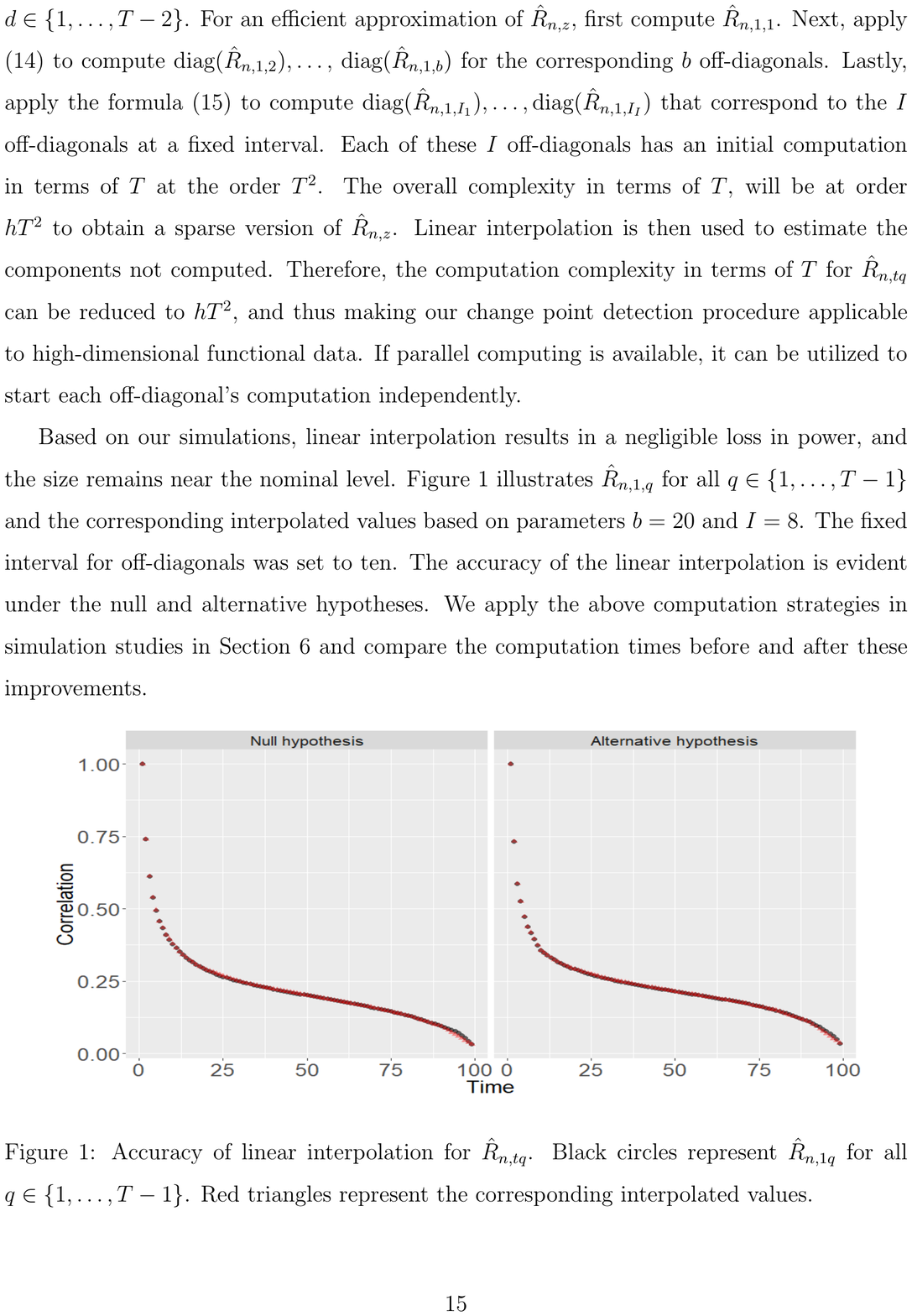}
\end{figure}

\begin{figure}[H]
  \includegraphics{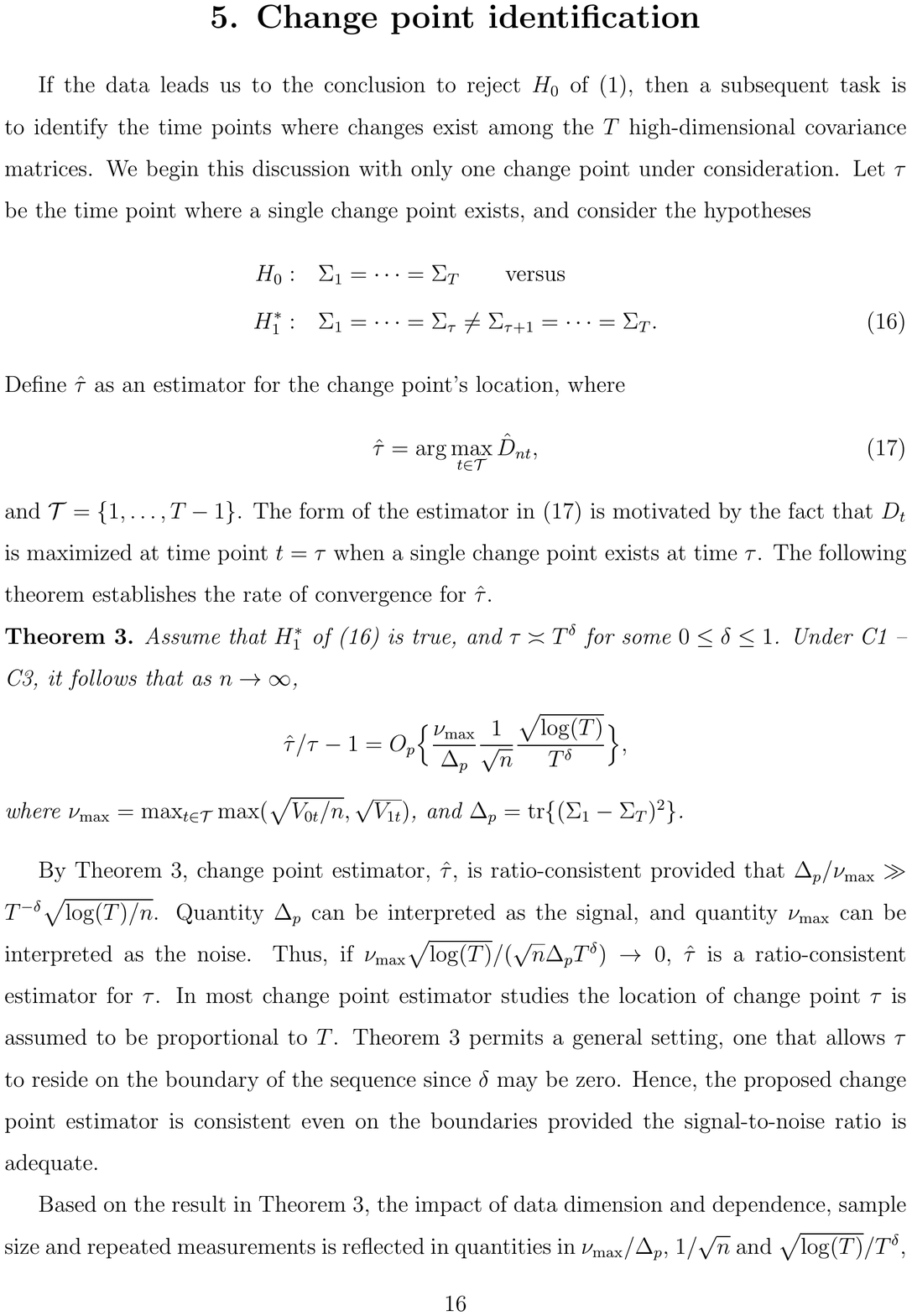}
\end{figure}

\begin{figure}[H]
  \includegraphics{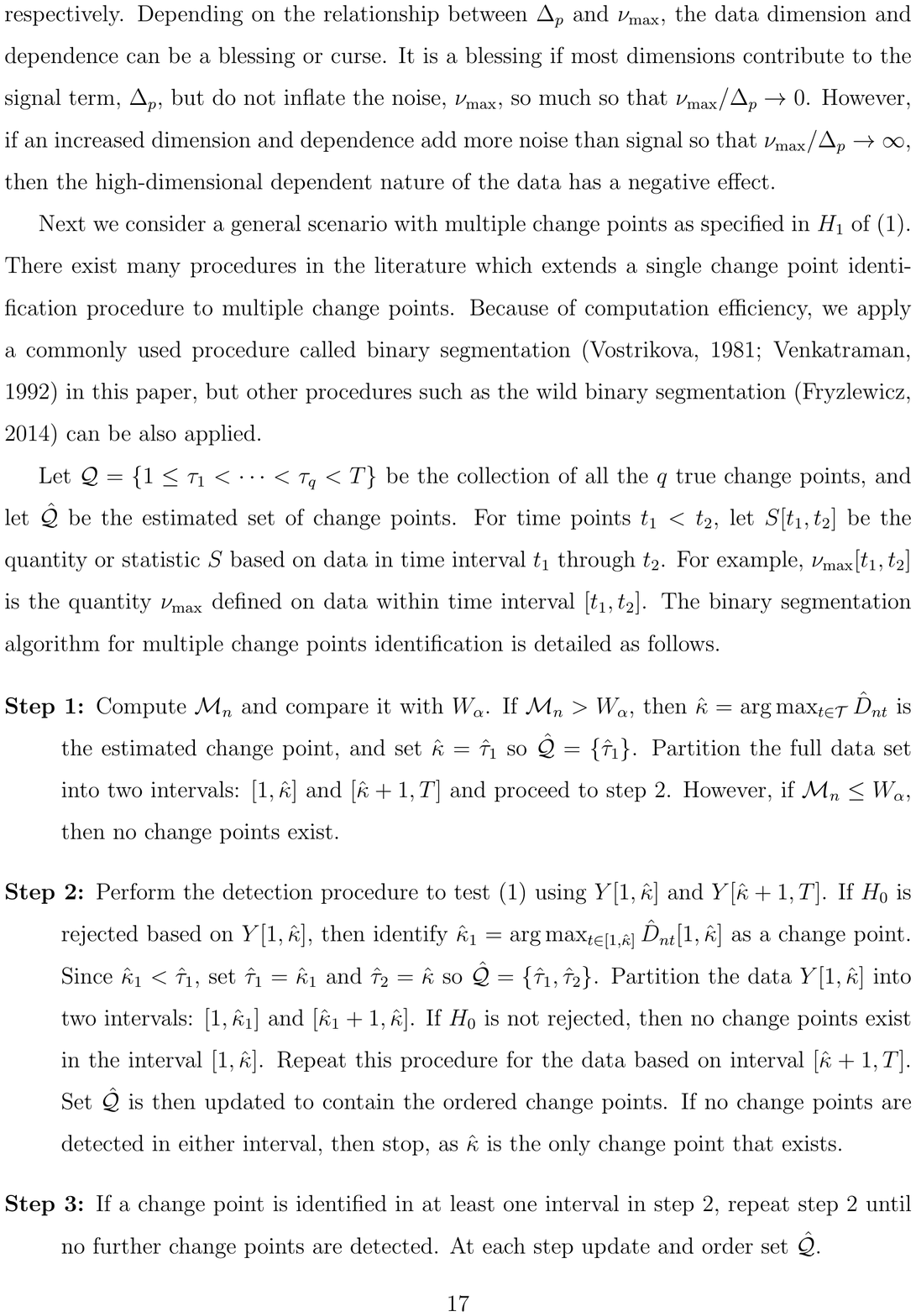}
\end{figure}

\begin{figure}[H]
  \includegraphics{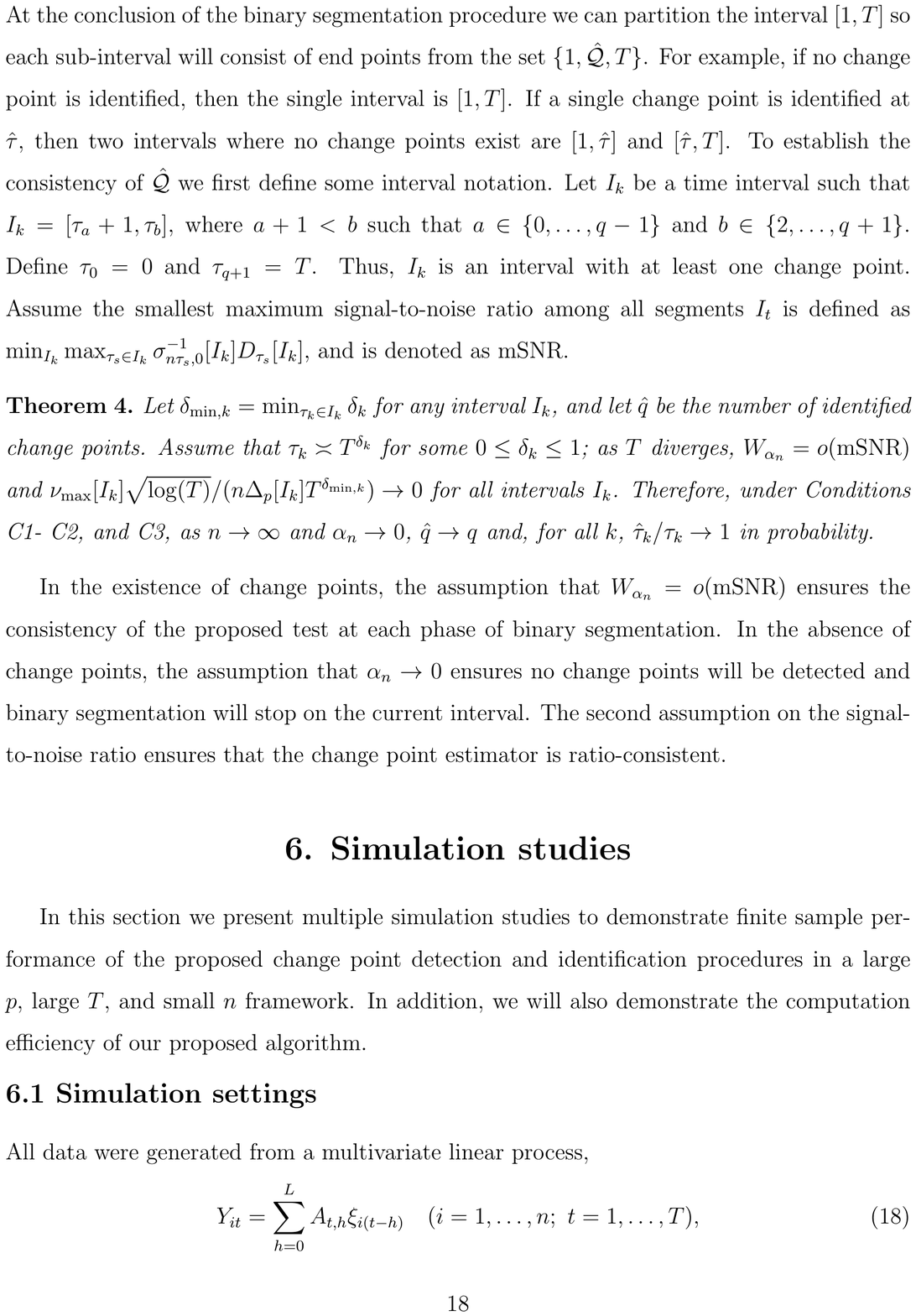}
\end{figure}

\clearpage
\restoregeometry
}

\noindent where  $A_{t,h}$ is a $p\times p$ matrix, and $\xi_{i(t - h)}$ are $p$-dimensional multivariate normally distributed random vectors with mean $0$ and covariance $I_p$. 
The data generation scheme given by (18) permits both spatial and temporal dependence. Spatial dependence occurs among the vector $Y_{it}$ for a given time point $t$,
and temporal dependence exists among $\{Y_{it}\}_{t=1}^T$ at different time points. By definition of $Y_{it}$ in (18),  
$\textup{cov}(Y_{it},Y_{is}) =\sum_{h=t-s}^{L} A_{t,h}A^\T_{s,h-(t-s)}I(t-s\leq L)$ if $t\geq s$. The spatial and temporal dependence is governed by the simulation parameter $L$.
%\begin{align*}
%	\textup{cov}(Y_{it},Y_{is}) = \left\{
%        \begin{array}{ll}
%            \displaystyle \sum_{h=t-s}^{L} A_{t,h}A^\T_{s,h-(t-s)}, & \quad t-s \leq L; \\
%            0, & \quad t-s > L.\\
%        \end{array}
%    \right.
%\end{align*}

For both change point detection and identification, we considered $n=40, 50$ and 60; $p=500, 750$ and 1000; and $T=50$ and 100.
We considered an additional case with $T = 150$ for change point identification. Parameter $L$ was set to be $3$.
Simulation results reported in Tables \ref{testing-large-T}, \ref{testing-large-T-linear-approx1} were based on 500 simulation replications, 
and simulation results in Table \ref{cpi-large-T} were based on 100 simulation replications. 

To evaluate the impact of dependence on the proposed procedures, we considered two structured types of matrices $A_{t,h}$ in order to control 
the spatial and temporal dependence. Exponential decay $A_{t,h}$ matrices were used for change point detection, and polynomial decay $A_{t,h}$ matrices were used for change point identification.

Define two matrices $B_1=\big\{(\mbox{0.6})^{|i-j|}I(|i-j|<p/5)\big\}$ and $B_2=\big\{(\mbox{0.6}+\delta)^{|i-j|}I(|i-j|<p/5)\big\}$,
where $(i, j)$ represents the $i$th row and $j$th column of the $p \times p$ matrices $B_1$ and $B_2$. 
Let $\tau_1=\floor{T/2}$ be the possible true underlying change point among the covariance matrices, where $\floor{x}$ is the floor function. For all $h \in \{0, \ldots, L\}$, define
$A_{t,h} =B_1$ for $t \in \{1, \ldots, \tau_1\}$, and $A_{t,h}=B_2$ for $t \in \{\tau_1 + 1, \ldots, T\}$.  Parameter $\delta$ in $B_2$ governs the signal strength in terms of 
how different the covariance matrices are before and after the change point at time $\tau_1$. When $\delta = 0$, then $B_1 = B_2$ and matrices $A_{t, h}$ are the same for all $t$; thus there is no
change point among covariance matrices, and the null hypothesis is true.
If $\delta > 0$, then the null hypothesis is false, and $\tau_1$ is the true covariance change point. 
For the change point detection simulation results in Table \ref{testing-large-T}, $\delta$ was set to $0.00, 0.025, 0.05$ and 0.10.

To evaluate the performance of the change point identification procedure and the proposed binary segmentation procedure, consider two change points: $\tau_1$ and $\tau_2$. 
Let $\tau_1 = \floor{T/2}$, and let $\tau_2 = \tau_1 + 2$. Define three matrices, $B_1^*=\big\{(\abs{i-j} + 1)^{-2}I(|i-j|<p/5)\big\}$, $B_2^*=\big\{(\abs{i-j} + \delta^* + 1)^{-2}I(|i-j|<p/5)\big\}$, 
and $B_3^*=\big\{(\abs{i-j} + 2\delta^* + 1)^{-2}I(|i-j|<p/5)\big\}$, where $(i, j)$ represents the $i$th row and $j$th column of the $p \times p$ matrices $B_1^*$, $B_2^*$, and $B_3^*$. 

\newpage 
\noindent Thus, for $h \in \{0, \ldots, 3\}$, $A_{t,h} = B_1^*$ for $t \in \{1, \ldots, \tau_1\}$, $A_{t,h}=B_2^*$ for $t \in \{\tau_1 + 1 \ldots, \tau_2\}$ and $A_{t,h}^*=B_3^*$ for $t \in \{\tau_2 + 1 \ldots, T\}$.
Since our purpose is to demonstrate the finite sample accuracy of change point identification, we chose values of $\delta$ to be 0.15, 0.25, and 0.35.

Two measures were considered to evaluate the change point identification procedure's efficacy: average true positives and average true negatives. For each simulation replication there exists two true change points at time 
points $\tau_1$ and $\tau_2$, and there exists $T-3$ time points where no change point exists. The average true positives are defined as the average number of correctly identified change points among 100 simulation 
replications. Similarly, the average true negatives are defined as the average number of correctly identified time points where no covariance change exists among 100 simulation replications.

\subsection*{6.2 Simulation results with exactly computed quantiles}

Table \ref{testing-large-T} demonstrates the empirical size and power of the proposed test procedure at nominal level $5\%$. The quantiles were obtained from the zero-mean 
multivariate Gaussian process, where every element of correlation matrix $\hat{R}_{n,tq}$ was computed exactly. We observe that the empirical size is well controlled for all combinations of $n$, $p$, and $T$. 
For a fixed $p$ and $T$, as $n$ increases the power increases. Likewise, as $\delta$ increases, the power of the change point detection procedure increases. 
For a fixed $n$ and $p$, the power also increases as $T$ increases. 

\begin{table}[bthp!]
\footnotesize 
\caption{Empirical size and power of the proposed change point detection test. The numbers in the table are percentages of simulation replications that reject the null hypothesis at nominal 
level $5\%$.}
\label{testing-large-T}
\centering
\begin{tabular}{cccrrrrrrrrr}\\
\hline 
& & & \multicolumn{3}{c}{$T=50$} & \multicolumn{3}{c}{$T=100$}\\
\cmidrule[0.4pt](lr{0.125em}){4-6} \cmidrule[0.4pt](lr{0.125em}){7-9} 
  &$\delta$        & $n/p$  & 500  & 750 & 1000   & 500  & 750 & 1000 \\
  \hline
  &                & 40   &  4.4 & 4.6 & 3.8          &  3.6   & 5.4   &   4.4\\
  & 0 (size)        & 50   &  4.8 & 4.0 & 3.6      &  2.0   & 4.6   &   4.0\\ 
  &                & 60   &  3.8 & 4.2 & 2.8          &  5.4   & 3.6   &   5.6\\ \cline{2-9}
  &                & 40   &  13.4 & 13.4 & 10.8    &  18.0 & 19.0 & 18.0\\
  & 0.025          & 50   & 17.0 & 19.2 & 17.0 &  30.6 & 27.2 & 30.4\\ 
  &                & 60   &  26.4 & 26.0 & 27.4    &  47.0 & 41.6 & 41.6\\ \cline{2-9}
  &                & 40   &  96.0 & 97.0 & 98.0    &  100  & 100  &  100\\
  & 0.05           & 50   &  100 & 100 & 100    &  100  & 100  &  100\\
  &                & 60   &  100 & 100 & 100       &  100  & 100  &  100\\ \cline{2-9}
  &                & 40   &  100 & 100 & 100       &  100  & 100  &  100\\
  & 0.10           & 50   &  100 & 100 & 100    &  100  & 100  &  100\\
  &                & 60   &  100 & 100 & 100       &  100  & 100  &  100\\
\hline 
\end{tabular}
\end{table}

Table \ref{cpi-large-T} provides the average true positives and average true negatives along with their corresponding standard errors 
of the proposed identification and binary segmentation procedure in the large $p$, large $T$, and small $n$ setting. 
For fixed $p$, $n$, and $T$, the average true positives and average true negatives approach their true values two and $T-3$, respectively, as $\delta$ increases. 
As the sample size increases, the average true positives and average true negatives approach their maximal values.  Our results indicate
that the proposed methods work well under various scenarios and are consistent in locating change points when the signal-to-noise ratio is
increased.

\begin{table}[bthp!]
\footnotesize
\caption{Average true positives (ATP) and average true negatives (ATN) for identifying multiple change points using the proposed binary segmentation method. 
%The maximum number of true positives for a given replication is 2. The maximum number of true negatives for a given replication is $T-3$. 
The standard errors of ATP and ATN are included in the parentheses. }
\label{cpi-large-T}
\centering
\begin{tabular}{ccccccccc}  \\
\hline  
& & &\multicolumn{2}{c}{$\delta$=0.15} & \multicolumn{2}{c}{$\delta$=0.25} & \multicolumn{2}{c}{$\delta$=0.35}\\ 
\cmidrule[0.4pt](lr{0.125em}){4-5} \cmidrule[0.4pt](lr{0.125em}){6-7} \cmidrule[0.4pt](lr{0.125em}){8-9} 
$T$ & $p$ & $n$ & ATP & ATN &  ATP & ATN &  ATP & ATN\\ 
\hline
\multirow{9}{*}{ 50 } & \multirow{3}{*}{ 500 } & 40 & 1.20 (0.40) &   46.76  (0.62) & 1.68  (0.47)   &   46.48  (0.52)   & 1.97   (0.17)  &   46.62  (0.49)   \\
                                                                  & & 50 & 1.41 (0.49) &   46.68  (0.53) & 1.91  (0.29)   &   46.42  (0.52)   & 2.00   (0.00)  &   46.63  (0.49)   \\
                                                                  & & 60 & 1.57 (0.50) &   46.58  (0.55) & 1.98  (0.14)   &   46.52  (0.50)   & 2.00   (0.00)  &   46.61  (0.55)   \\ \cline{2-9}
& \multirow{3}{*}{ 750 }                                 & 40 & 1.30 (0.46) &   46.78  (0.42) & 1.77  (0.42)   &   46.51  (0.50)   & 2.00   (0.00)  &   46.59  (0.61)   \\
                                                                  & & 50 & 1.33 (0.47) &   46.66  (0.48) & 1.95  (0.22)   &   46.53  (0.50)   & 2.00   (0.00)  &   46.70  (0.46)   \\
                                                                  & & 60 & 1.57 (0.50) &   46.58  (0.55) & 1.99  (0.10)   &   46.53  (0.56)   & 2.00   (0.00)  &   46.64  (0.50)   \\ \cline{2-9}
& \multirow{3}{*}{ 1000 }                               & 40 & 1.27 (0.45) &   46.76  (0.55) & 1.81  (0.39)   &   46.61  (0.51)   & 1.95   (0.22)  &   46.59  (0.50)   \\
                                                                  & & 50 & 1.48 (0.50) &   46.67  (0.47) & 1.95  (0.22)   &   46.58  (0.50)   & 2.00   (0.00)  &   46.76  (0.43)   \\
                                                                  & & 60 & 1.65 (0.48) &   46.51  (0.63) & 1.99  (0.10)   &   46.69  (0.47)   & 2.00   (0.00)  &   46.59  (0.67)   \\ \cline{1-9}
\multirow{9}{*}{ 100 } &\multirow{3}{*}{ 500} & 40 & 1.27 (0.45) &   96.75  (0.50) & 1.74  (0.44)   &   96.56  (0.50)   & 1.98   (0.14)  &   96.54  (0.52)   \\
                                                                  & & 50 & 1.31 (0.47) &   96.67  (0.47) & 1.92  (0.27)   &   96.54  (0.50)   & 2.00   (0.00)  &   96.44  (0.50)   \\
                                                                  & & 60 & 1.62 (0.49) &   96.70  (0.48) & 1.99  (0.10)   &   96.56  (0.52)   & 2.00   (0.00)  &   96.46  (0.54)   \\ \cline{2-9}
& \multirow{3}{*}{ 750 }                                 & 40 & 1.22 (0.42) &   96.76  (0.50) & 1.85  (0.36)   &   96.59  (0.51)   & 1.98   (0.14)  &   96.54  (0.50)   \\
                                                                  & & 50 & 1.33 (0.47) &   96.59  (0.55) & 1.96  (0.20)   &   96.51  (0.50)   & 2.00   (0.00)  &   96.54  (0.50)   \\
                                                                  & & 60 & 1.60 (0.49) &   96.59  (0.49) & 1.99  (0.10)   &   96.55  (0.50)   & 2.00   (0.00)  &   96.42  (0.78)   \\ \cline{2-9}
& \multirow{3}{*}{ 1000 }                               & 40 & 1.20 (0.40) &   96.80  (0.43) & 1.74  (0.44)   &   96.52  (0.50)   & 1.98   (0.14)  &   96.59  (0.55)   \\
                                                                  & & 50 & 1.34 (0.48) &   96.64  (0.50) & 1.90  (0.30)   &   96.50  (0.52)   & 2.00   (0.00)  &   96.49  (0.50)   \\
                                                                  & & 60 & 1.59 (0.49) &   96.50  (0.61) & 2.00  (0.00)   &   96.58  (0.50)   & 2.00   (0.00)  &   96.44  (0.61)   \\ \cline{1-9}
\multirow{9}{*}{ 150 } &\multirow{3}{*}{ 500} & 40 & 1.19 (0.39) & 146.76  (0.43) & 1.73  (0.45)   & 146.53  (0.56)   & 1.97   (0.17)  & 146.48  (0.56)   \\
                                                                  & & 50 & 1.34 (0.48) & 146.68  (0.47) & 1.95  (0.22)   & 146.55  (0.50)   & 2.00   (0.00)  & 146.40  (0.53)   \\
                                                                  & & 60 & 1.54 (0.50) & 146.51  (0.52) & 2.00  (0.00)   & 146.57  (0.50)   & 2.00   (0.00)  & 146.53  (0.52)   \\ \cline{2-9}
& \multirow{3}{*}{ 750 }                                 & 40 & 1.16 (0.37) & 146.84  (0.40) & 1.73  (0.45)   & 146.58  (0.50)   & 1.97   (0.17)  & 146.46  (0.58)   \\
                                                                  & & 50 & 1.42 (0.50) & 146.64  (0.50) & 1.97  (0.17)   & 146.55  (0.52)   & 2.00   (0.00)  & 146.52  (0.56)   \\
                                                                  & & 60 & 1.56 (0.50) & 146.45  (0.58) & 1.98  (0.14)   & 146.42  (0.52)   & 2.00   (0.00)  & 146.55  (0.56)   \\ \cline{2-9}
& \multirow{3}{*}{ 1000 }                               & 40 & 1.20 (0.40) & 146.80  (0.40) & 1.72  (0.45)   & 146.49  (0.50)   & 1.97   (0.18)  & 146.51  (0.52)   \\
                                                                  & & 50 & 1.46 (0.50) & 146.70  (0.46) & 1.92  (0.27)   & 146.50  (0.67)   & 2.00   (0.00)  & 146.47  (0.63)   \\
                                                                  & & 60 & 1.53 (0.50) & 146.56  (0.54) & 1.99  (0.10)   & 146.56  (0.52)   & 2.00   (0.00)  & 146.51  (0.50)   \\
\hline
\end{tabular}
\end{table}

\subsection*{6.3 Approximation algorithm and improved computation time}

To illustrate the accuracy of the proposed approximation algorithm in Section 4, Table \ref{testing-large-T-linear-approx1} conveys the empirical size and power of the proposed test procedure.
 Rather than compute $\hat{R}_{n,tq}$ for all $t, q \in \{1, \ldots, T - 1\}$, we computed the first $b$ off-diagonals and the last $w$ columns of $\hat{R}_{n,tq}$. 
The remaining values were imputed via linear interpolation.  In the simulation studies, parameter $b = 5$, and $w = 5$, 10 and 20. We observe that
the sizes of the test based on the approximated quantiles are well maintained at the nominal level of 0.05, and there are minuscule differences from the empirical 
sizes in Table \ref{testing-large-T}. Furthermore, there is only a minimal difference in power when compared to the corresponding results in Table \ref{testing-large-T}. 
These results indicate the proposed approximation maintains accuracy. The simulation results for cases when $b=10$ and 20 are included in the supplementary files.

\begin{table}[htbp!]
\footnotesize 
\caption{Empirical size and power of the proposed test for $T=100$, percentages of simulation replications that reject the null hypothesis, quantile computed from a correlation matrix that used proposed approximation in Section 4. The first 5 off-diagonals were computed exactly as well as the last $w$ components for each row}
\label{testing-large-T-linear-approx1}
\centering
\begin{tabular}{cccrrrrrrrrr}   
\\
\hline 
& & & \multicolumn{3}{c}{$w=5$} & \multicolumn{3}{c}{$w=10$} & \multicolumn{3}{c}{$w=20$} \\ 
\cmidrule[0.4pt](lr{0.125em}){4-6} \cmidrule[0.4pt](lr{0.125em}){7-9} \cmidrule[0.4pt](lr{0.125em}){10-12}
  &$\delta$        & $n/p$  & 500  & 750 & 1000   & 500  & 750 & 1000 & 500  & 750 & 1000 \\
  \hline
  &                & 40   &  $3.4$ & $4.8$ & $4.2$ & $3.4$ & $4.8$ & $4.2$ & $3.4$ & $5.2$ & $4.2$ \\
  & 0(size)        & 50   &  $2.0$ & $4.6$ & $3.8$ & $2.0$ & $4.6$ & $4.0$ & $2.0$ & $4.6$ & $4.0$ \\
  &                & 60   &  $4.8$ & $3.2$ & $5.0$ & $4.8$ & $3.2$ & $5.0$ & $5.2$ & $3.8$ & $5.6$ \\ \cline{2-12}
  &                & 40   &  $17.8$ & $19.0$ & $17.6$ & $17.8$ & $19.0$ & $17.6$ & $17.8$ & $19.0$ & $17.6$ \\
  & 0.025          & 50   &  $30.8$ & $26.2$ & $30.2$ & $30.8$ & $26.6$ & $30.2$ & $30.8$ & $26.6$ & $30.2$ \\ 
  &                & 60   &  $46.6$ & $40.8$ & $41.0$ & $46.6$ & $40.8$ & $41.0$ & $46.6$ & $41.2$ & $41.0$ \\  \cline{2-12}
  &                & 40   &  $100$ & $100$ & $100$ & $100$ & $100$ & $100$ &  100 & 100 & 100\\
  & 0.050           & 50   &  100 & 100 & 100    &  100 & 100 & 100 &  100 & 100 & 100\\
  &                & 60   &  100 & 100 & 100    &  100 & 100 & 100 &  100 & 100 & 100\\ \cline{2-12}
  &                & 40   &  100 & 100 & 100    &  100 & 100 & 100 &  100 & 100 & 100\\
  & 0.100           & 50   &  100 & 100 & 100    &  100 & 100 & 100 &  100 & 100 & 100\\
  &                & 60   &  100 & 100 & 100    &  100 & 100 & 100 &  100 & 100 & 100\\
  \hline 
\end{tabular}
\end{table}

%The computation complexity reduced significantly after applying the proposed approximation algorithm. 
Applying expressions (11) -- (14) we evaluate and compare the computation time with that of the naive detection method.  
Our algorithm improves the computation efficiency with respect to the sample size $n$ and the number
of repeated measurements $T$. To see the improvements, we isolate these two parameters and consider two cases. 
In the first case, we consider $p=1, n=4$ and $T=50, 75, 100, 125, 150$ so so we can
quantify the affect of reducing the computation complexity with
respect to $T$. In the second case, we consider $p=1, T=2$ and $n=30, 60, 90, 120, 150$
so that we can evaluate the saved computation time with respect to $n$. 
Figure \ref{comptime} compares the median computation time of the naive method and the proposed method using expressions (11) -- (14).
Each parameter combination was run for 100 simulation replications.
The left plot illustrates the computation time in seconds as
$T$ varies for each method. When $T=150$, $n=4$ and $p=1$, the
naive method had a median run time of 36830.685 seconds; while for the
same data and parameter combinations the proposed method had a median run time of 429.898 seconds. This is equivalent to 10.23 hours versus
7.16 minutes. Similarly, the right plot illustrates the computation
time in seconds as $n$ varies for each method. For $n=150$, $T=2$ and
$p=1$ the median run time of the proposed method is over 266 times faster than the median run time of the naive method.

\setcounter{figure}{1}
\begin{figure}[htbp!]
  \centering
  \includegraphics[scale=.42]{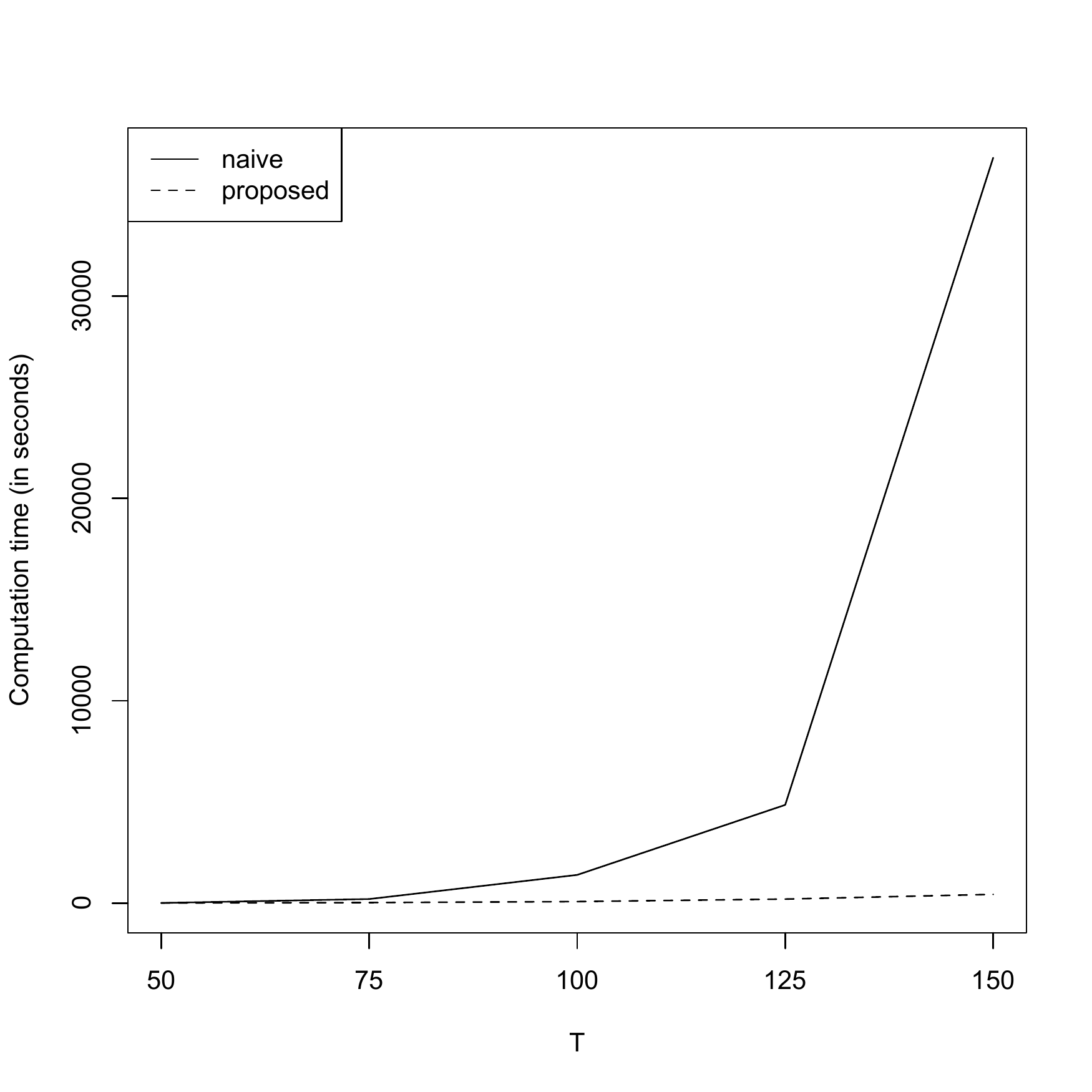}
  \includegraphics[scale=.42]{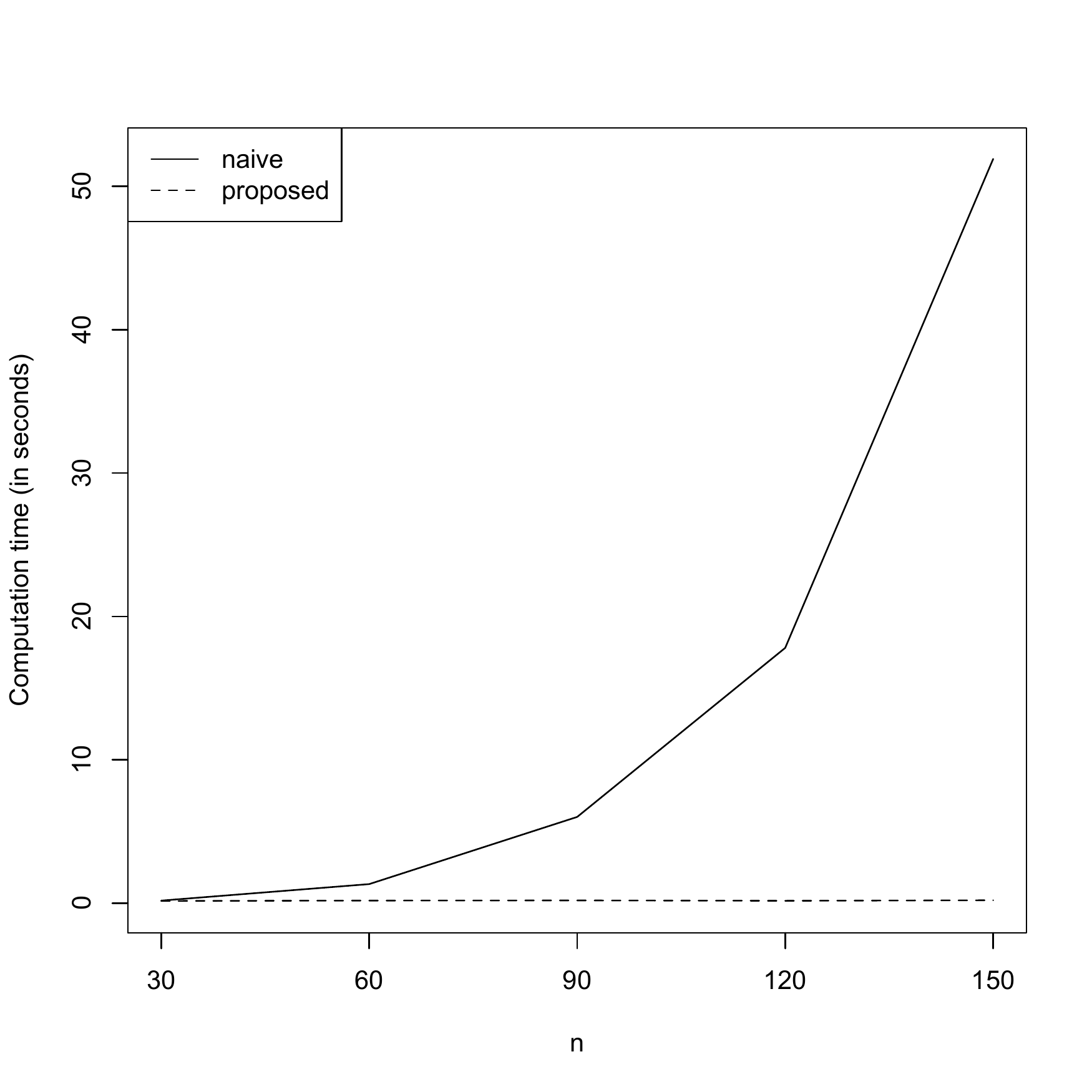}
  \caption{Computation time of the naive method and proposed computation method in Section 4. The left panel is the comparison with respect to $T$ while
  the right panel is with respect to $n$.}
  \label{comptime}
\end{figure}

\setcounter{section}{7}
{\centering{
\section*{7. An application to event segmentation}
\label{sec:an-empirical-study}
}}
%Human memory has been studied through fMRI experiments in the context of discrete and continuous activities. 
%One goal of neurologists is to better understand perception and memory processes in humans as they experience continuous real-world events (Baldassano et al. 2017). 
Event segmentation theory (Zacks et al., 2007) suggests that humans generate event boundaries in memory by partitioning a continuous experience into a series of segmented discrete events.  
Schapiro et al. (2013) discovered that event boundaries are formed around changes in functional connectivity. In this section we apply our
proposed procedure to detect and identify change points in covariance matrices in the fMRI data set collected by Chen et al. (2017) to partition fMRI data into a series of segments with 
static functional connectivity within each segment. Time points where functional connectivity changes may represent event boundaries as suggested in 
the aforementioned neuroscience literature.

We apply our proposed method to the motivating task-based fMRI experiment (Chen et al., 2017) introduced in Section 1. The experiment involved 17 
participants that each watched the same 48-minute segment (episode 1: ``a study of pink'') of the BBC television series {\it Sherlock} while undergoing an fMRI scan. 
None of the participants had watched the series {\it Sherlock} prior to the study. A 30-second cartoon was prepended to the movie to allow the brain time to adjust to new audio and visual stimuli. 
Including an unrelated cartoon prior to studies such as this is common practice as it reduces statistical noise. Subjects were instructed to watch the 
television episode as they would watch a typical television episode in their own home. Data were gathered from a Siemens Skyra 3T full-body scanner, and
the fMRI machine acquired an image of each participant's brain every 1.5 seconds. As a result, the 48-minute segment of {\it Sherlock} resulted in 1,976 repeated measurements.
More details about the experiment and processes of acquiring functional and anatomical images are provided in Chen et al. (2017). 

\begin{figure}[hbtp!]
  \centering
  \includegraphics[scale=.65]{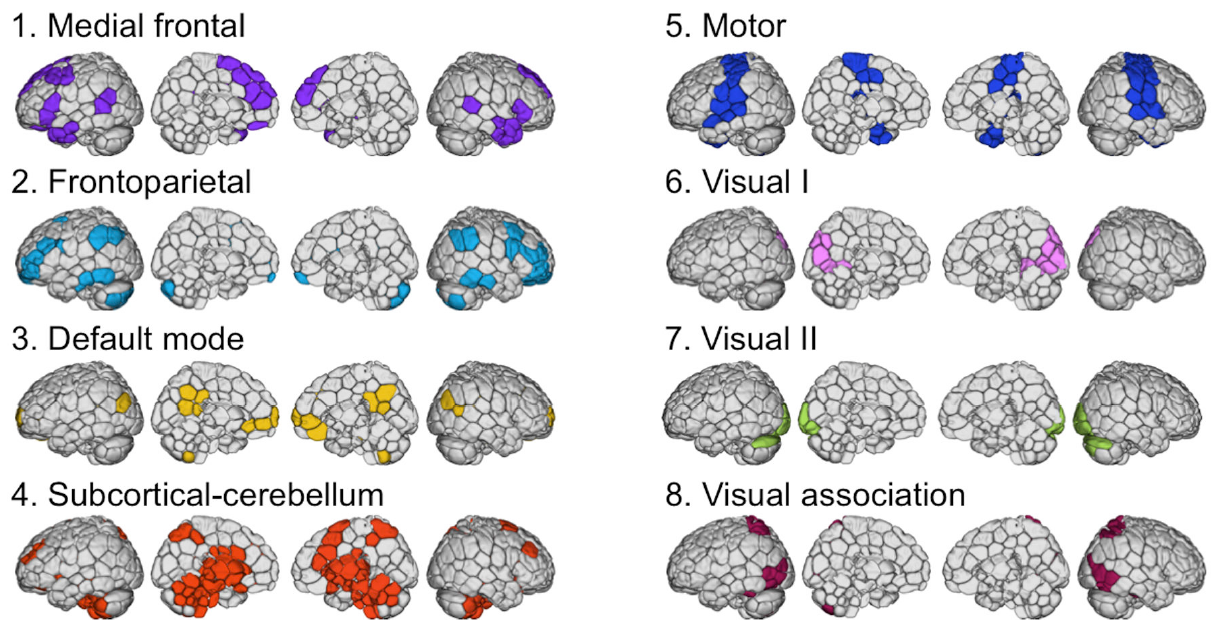}
  \caption{268 Shen-node parcellation. This image was obtained from Finn et al. (2015).}
  \label{shen-regions}
\end{figure}

To demonstrate our proposed methods, we analyzed the first 131 time points of fMRI data which corresponds to the preface of {\it Sherlock}, 
or equivalently the first three minutes and 16 seconds of the movie. Let $Y_{it}$ $(i =1, \dots, 17; \ t = 1, \dots, 131)$ be the 268-dimensional 
random vector containing the BOLD measurements for the 268 nodes of the $i$th individual at time $t$. A node, or region of interest, represents a 
collection of voxels. The 268 node parcellation was performed according to Shen et al. (2013), where voxel groupings ensure functional homogeneity within each node, making it ideal for node network and dynamic 
functional connectivity analysis. Figure \ref{shen-regions} illustrates the 268 Shen node parcellation along with large-scale node groupings. 
A node-level analysis, as opposed to a voxel-level analysis, allows for more interpretable results.
For further details on the benefits and processes of Shen node parcellation, we refer readers to Shen et al. (2013). 
In summary, the fMRI data contains $p = 268$ random variables that are repeatedly
measured for $T = 131$ times from $n = 17$ subjects. 

Let $\Sigma_t$ be the covariance matrix of $Y_{it}$ ($t=1,\cdots, 131$). Matrix $\Sigma_t$, or its inverse, is a $268\times 268$-dim matrix that represents the functional connectivity among 268 nodes at time $t$.  
The goal of our analysis is to detect if there is any changes among $\Sigma_t$'s. If change point exists, we further identify the locations of the change points, 
which will be applied to partition the movie preface into discrete event segments. We first applied our test procedure proposed in Section 3 and obtained the test statistic value $\mathcal{M}_n = 4.433$; 
we rejected $H_0$ of (\ref{hyp-infinite-T}) as the p-value was less than $0.001$. This result suggests that the functional connectivity among the 268 nodes
changes over time. Accordingly, we applied the proposed binary segmentation to identify all significant change points among $131$ covariance matrices. Our proposed method identified 20 change points at time points
2,  25,  36,  39,  40,  41,  42,  58,  60,  61,  63,  81,  83, 110, 113, 114, 115, 116, 128 and 130. Based on the identified change points, we observe that several clusters exist. 
The first cluster occurred at 39, 40, 41 and 42, the second cluster occurred at 58, 60, 61 and 63, the third cluster occurred at 
81 and 83, and the last cluster appeared at 110, 113, 114, 115 and 116. This phenomena indicates that the functional connectivity changes slowly, rather than abruptly, around these change points, 
and different individuals may have slightly different points of change due to individual heterogeneity. 

To use these change points for event segmentation, we group the sets of change points that are close to each other because these sets of change points essentially imply the same event boundaries. 
After grouping, the change points used for event segmentation are 2, 25, 36, 42, 58, 83, 113 and 130, where 42, 58, 83 and 113 are representations of their respective group
of event boundaries. Change points 2 and 130 coincide with the beginning of the stimuli and the end of preface. Therefore, change points 25, 36, 42, 58, 83 and 113 are used for event segmentation, 
which corresponds to 38, 54, 63, 87, 125 and 170 seconds in the movie.

\begin{figure}[htbp!]
  \centering
  \includegraphics[scale=.5]{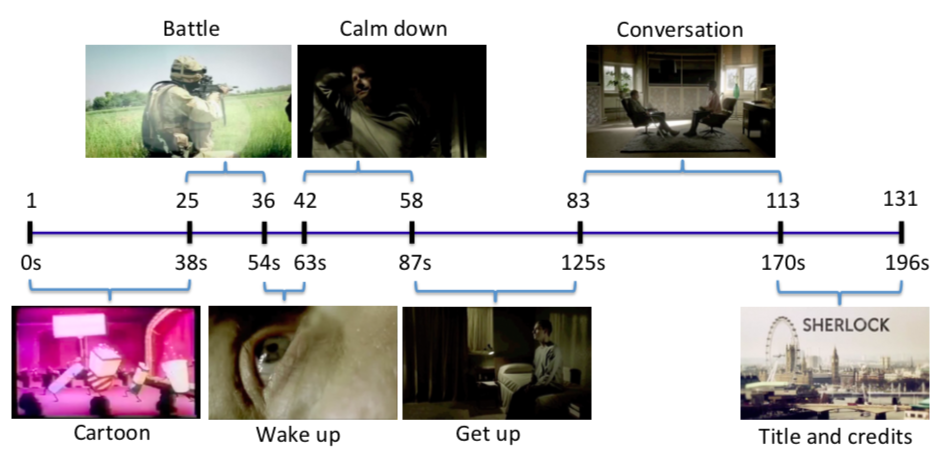}
\caption{Event segmentation based on the covariance change points in fMRI data detected and identified by using the proposed approaches. Images were obtained from the movie
{\it Sherlock}.}
\label{filmsegments}
\end{figure}

\begin{figure}[htbp!]
\centering
\includegraphics[height=0.7\textheight, width=0.95\textwidth]{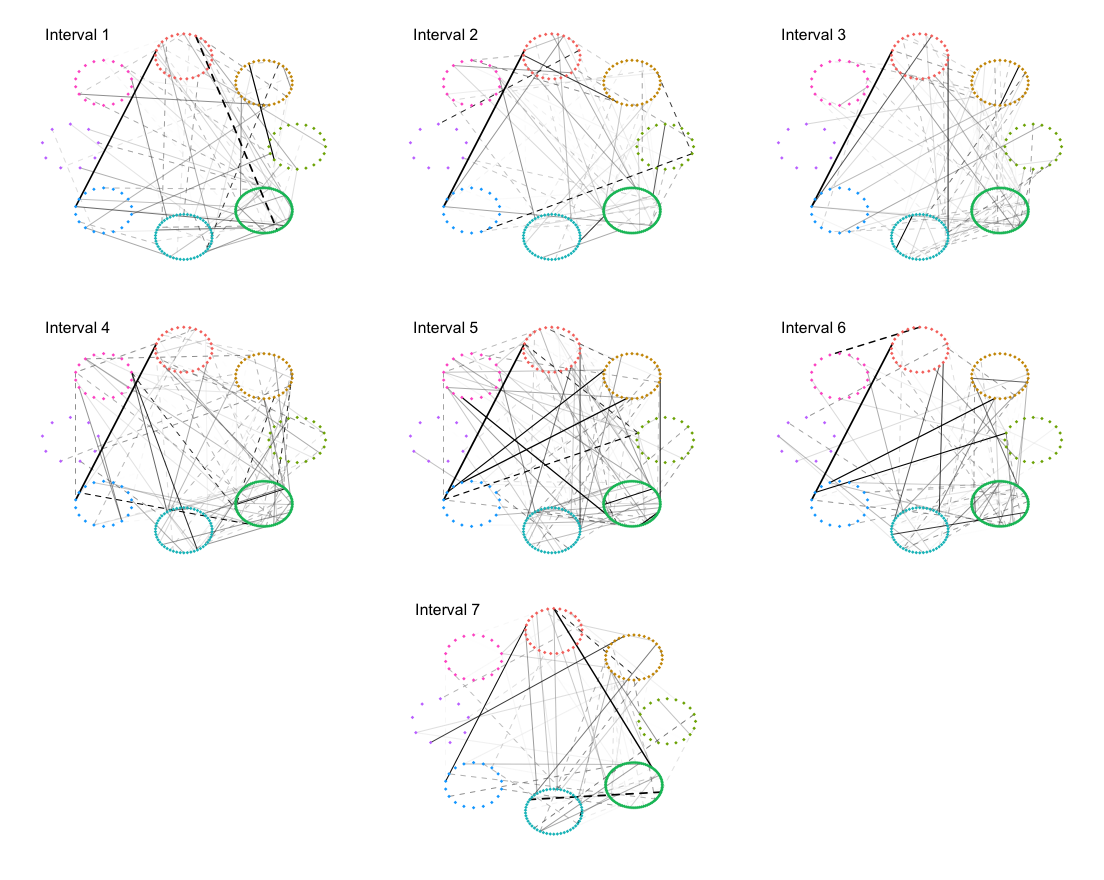}
\caption{Correlation networks based on an average over a time interval in which the covariance matrices are homogeneous. Each circle is comprised of 67 Shen nodes. Solid lines represent a positive correlation, and dashed lines represent a negative correlation. The darker the line the stronger the correlation between nodes. A correlation threshold value of 0.70 in absolute values was used.}
\label{fmri-network-plot-1}
\end{figure}

Figure \ref{filmsegments} illustrates the final event segmentation using the proposed approach and the corresponding events in each segment.
The event partitions based on the identified change points coincide with important situations in the television episode {\it Sherlock}. The first segment is from the beginning to the first change point (38s in the movie), 
which corresponds to the cartoon ``let us all go to the lobby'' prepended to the movie. The second segment, between 38s and 54s, in the movie corresponds to flashbacks of battle scenes. The third segment, between 54s and 63s, corresponds to the period of Watson, the main character, awakening from his dream. 
The fourth segment, between 63s and 87s, is a period where Watson calms down. Between 87 and 125 seconds is the period that Watson got up 
and prepared to write his blog. 
During the sixth segment, between 125s and 170s, Watson is having a conversation with a journalist. The final segment, from 170s to the end of the preface, corresponds to the title and introductory credits.

Figures \ref{fmri-network-plot-1} demonstrates the changes in covariance matrices around the change points used in Figure \ref{filmsegments}. Each subplot is the estimated correlation matrix between nodes using the data 
between change points. For example, in Figure \ref{fmri-network-plot-1}, the correlation network in interval 1 is estimated based fMRI data between time interval $[3, 25]$.  
We observe that the correlation networks estimated from different intervals are significantly different from each other. This indicates that the identified change points are consistent with the changes in correlation networks. 
Correlation network layouts are structured according to the eight large-scale node groupings illustrated in Figure \ref{shen-regions}. 
The top-centered circle consists of nodes within the medial frontal group. Moving clockwise on a given sub-plot, the remaining circles represent frontoparietal, default mode, 
subcortical-cerebellum, motor, visual I, visual II, and visual association.

\end{document}